\documentclass[runningheads]{llncs}
\usepackage{graphicx}
\usepackage{multicol}
\begin{document}
\title{Is this IoT Device Likely to be Secure? \\ Risk Score Prediction for IoT Devices Using Gradient Boosting Machines}
%
%
\author{Carlos A. Rivera A.\inst{1} \and
Arash Shaghaghi\inst{2, 1} \and
David D. Nguyen\inst{1} \and
Salil S. Kanhere\inst{1}}
\authorrunning{C.A. Rivera A. et al.}
%
\institute{
The University of New South Wales (UNSW), Sydney, Australia
\and
RMIT University, Melbourne, Australia \\
\email{\{c.riveraalvarez, d.d.nguyen, salil.kanhere\}@unsw.edu.au} \\
\email{arash.shaghaghi@rmit.edu.au}
}
\titlerunning{Is this IoT Device Likely to be Secure?} 
\maketitle              
\begin{abstract}    
Security risk assessment and prediction are critical for organisations deploying Internet of Things (IoT) devices.  An absolute minimum requirement for enterprises is to verify the security risk of IoT devices for the reported vulnerabilities in the National Vulnerability Database (NVD). This paper proposes a novel risk prediction for IoT devices based on publicly available information about them. Our solution provides an easy and cost-efficient solution for enterprises of all sizes to predict the security risk of deploying new IoT devices. After an extensive analysis of the NVD records over the past eight years, we have created a unique, systematic, and balanced dataset for vulnerable IoT devices, including key technical features complemented with functional and descriptive features available from public resources. We then use machine learning classification models such as Gradient Boosting Decision Trees (GBDT) over this dataset and achieve 71\% prediction accuracy in classifying the severity of device vulnerability score.

\end{abstract}

{\small
}

\section{Introduction}\label{section1}
Internet of Things (IoT) devices facilitate advancements and financial benefits for organisations and continue to transform and impact organisational processes and structures. Nonetheless, they expand the plethora of attacks already aimed at organisations. IoT devices present unique challenges that cannot be adequately managed using conventional approaches. Traditional security mechanisms and risk assessment frameworks deployed in organisations fall short when it comes to IoT devices \cite{nurse}. Moreover, decision-makers and technical staff responsible have been reported to overlook the security risks imposed by IoT devices due to their limited knowledge about them \cite{nurse}. \\

Typically, IoT devices are produced by manufacturers and sold `as is' \cite{arash}. The lack of standards and legal frameworks leave the security of devices as an option to manufacturers who are predominantly focused on issues such as cost, size, usability, and time-to-market. Vulnerability assessment of IoT devices is also costly for enterprises, given the exponential growth in numbers, variety, and configurations. Moreover, ongoing advancements in IoT devices limit the support period by manufacturers providing security patches \cite{eduardo}. As a result, IoT devices become `orphans' much earlier than traditional computing devices. Nonetheless, these vulnerable IoT have been reported to remain connected to networks leading to massive global security threats such as Mirai \cite{mirai}. In another example, a recent report by CyberCX confirmed that password-less IoT devices are still affecting various organisations \cite{cyberx}.   \\

This paper provides a risk score prediction solution for enterprises, allowing system administrators to carry out an easy and cost-efficient risk prediction based on an IoT device's publicly available information. We studied the National Vulnerability Database (NVD)\footnote{\url{https://nvd.nist.gov}} records and analysed 1153 vulnerable IoT devices reported within the past eight years. We then created a dataset of these vulnerable devices and collected a set of twenty-seven features for each device (e.g., manufacturer, device type, price, and authentication capability). Thereafter, a set of eleven features were shortlisted that allowed creating a balanced dataset. Using Gradient Boosting Decision Trees (GBDT) machine learning model over this dataset, we achieve 71\% accuracy in predicting the device vulnerability risk. Our solution is complementary to the vulnerability analysis of IoT devices (e.g., \cite{nacm,Meidan,Zhang}), which is an expensive and time-consuming process requiring in-depth analysis of devices (e.g., white-box analysis). In other words, our risk prediction solution allows filtering out high-risk IoT devices and reducing vulnerability analysis related costs. Risk prediction is also complementary to other IoT threat detection and prevention solutions actively explored in the IoT security literature (e.g., \cite{Skowron,Hasan,Kumar,Zeadally,Chin-Wei,Saha}). \\

The rest of this paper is structured as follows: we review the relevant literature in \S\ref{section2-RW}. In \S\ref{section3}, we  discuss in detail our research methodology, adversary model, assumptions, data transformation, the IoT vulnerability database created, and machine learning classification model used. In \S\ref{section4_Experiments}, we present our evaluation results and conclude the paper with a discussion and conclusions in \S\ref{section5_Discussion}.\par 

\section{Related Work}\label{section2-RW}
Automatic analysis and classification of vulnerability databases has been the subject of prior research (not specific to IoT). For instance, authors in \cite{Neuhaus}, used Topic Models to analyse security trends in CVE databases with no expert knowledge. In \cite{wang,nacm}, Bayesian Networks were used to categorise vulnerabilities from CVE databases according to their security types. To the best of our knowledge, however, risk prediction for IoT devices using their publicly accessible features is novel in the related literature. 

We believe the solution proposed by authors in \cite{Grzegorz} is the closest work to ours, where authors aim to predict the category of a vulnerable IoT device. Authors in \cite{Grzegorz} propose a classification of device-related vulnerability data for IoT and IIoT equipment. Authors divided Common Vulnerabilities and Exposures (CVE) records from NVD's public dataset into seven categories: home equipment, SCADA, etc. They then used the support vector machine (SVM) classifier on the device and vulnerability data to predict `new vulnerabilities' categories. To narrow the data universe, they opted only to obtain data flagged with the value h (hardware). 

In our proposal, we combine features from NVD with publicly accessible information about devices to create a new balanced (i.e., without missing data) dataset of vulnerable IoT devices. In other words, we noticed data imbalance based on information available in the paper. Specifically, this issue can be observed in one of the results presented, where the training data contained vulnerabilities from the years 2014 to 2017 and test data from the year 2018. Here, the classification model returned zero percentage in Recall and Precision for one of the classes and poor (below 55\%) percentages in precision in five other classes, leaving one class with results over 60\%. Whereas for Recall, only two classes had results above 60\%, leaving the rest with values below 45\%. These results indicate that the model used in \cite{Grzegorz} struggled to differentiate some of the classes. In our solution, we performed several processes to our dataset to determine the level of imbalance which was later leveraged over the model. We verified the effect of these corrective actions over the evaluation metrics and verified that the class imbalance was indeed lesser. Also, we confirm that the model increased the level of confidence predicting all the classes. 

Authors in \cite{Grzegorz} performed analysis over models rather than over the data. Whereas in our implementation, we analysed the data thoroughly to identify the adequate machine learning model. Although authors in \cite{Grzegorz} aim to predict the category of vulnerable IoT/IIOT devices, they used the entire dataset since the NVD dataset records contain poor references to their target type of devices. In our proposal, however, we narrow the analysis to only IoT devices, and we ensured that all the other features were directly related to the target devices. We also create a unique IoT vulnerability dataset with information about each device that could be re-used in future research. Finally, authors in \cite{Grzegorz} aimed to use data from previous years, e.g. 2015, as training data to predict new vulnerabilities from the year 2016 (as test data). Our model, however, contains records from the years 2013 to 2021 (up to 17th of June), and we split the data using the whole dataset rather than per each year. 

\section{Inception of Risk Score Prediction through Ensemble Models.}\label{section3}

\subsection{Dataset, Data Transformation, Models and Evaluation Metrics}\label{sub-section3.2_Models}

In this paper, we propose a classification model to predict the risk score (CVSS-based) of IoT devices. For this, we use create a unique IoT vulnerability inclusive of key technical features complemented with functional and descriptive features extracted from public resources. In the following sub-sections, we explain the structure of the dataset, collection method, transformation, models and evaluation metrics. \\

We make the following assumptions for the data to be of use: 
\renewcommand{\theenumi}{\Roman{enumi}}
\begin{enumerate}
\item We selected the NVD as the primary source of information about vulnerabilities related to IoT devices as well as different Internet sources to complement the records related to specific devices.  
\item The NVD database contains information about vulnerabilities and risk on many systems, however, we narrowed our scope of study to IoT devices in part of the following categories: Home security, Home Telecommunications, Small Office Telecommunication, Medical Smart Devices, Wearable Technology and Other. This last classification is used to describe devices such as Toys, Tracking devices, Accessories and Smart Storage. This classification is flexible so it can host a device that cannot be put in any other.
\item We acknowledge the existence of other types of IoT devices such as industrial controls, modules, controllers, telecommunications for middle or big organisations, mobile phones, tablets, and energy transmission. Though, none of them is included in the study. We, however, recognise the intention of the study is flexible and can be re-directed or extended.
\item The aim of the study does not require experimentation over any physical or virtual device(s).
\item The information obtained from the NVD contains records of IoT devices from the year 2013 and up to the current year, updated to the June version.
\end{enumerate}

\subsubsection{Dataset}\label{subsub-section_Dataset}
To prove the effectiveness and novel approach of our theory, we searched for datasets with content related to vulnerabilities and other information related to IoT devices. For it, we used the vulnerabilities database (NVD) of NIST (National Institute of Standards and Technology). This database is constructed from the CVE database and complemented with the vulnerability score (CVSS),  the technical specification of the affected software or hardware (Common Platform Enumeration - CPE), and a list of weaknesses related to the vulnerability disclosed (Common Weakness Enumeration - CWE). Note the CVE database contains the description of the vulnerability and affected product, a record id, references to web pages with detailed information of the vulnerability, and Date of record creation. We analysed the NVD and extracted information from the different elements to conform to an initial set of features. From this process and after processing several records we encounter several inconsistencies, for instance, the features expecting to receive information of devices had inconsistencies or were difficult to standardise. Because of these limitations, we concluded that using the NVD as the sole information provider was not ideal and decided to design a new dataset. This new dataset would use different sources, thus new features. This would resolve the limitations presented before. Note that the NVD would serve as the basis of our dataset but also have added information about IoT devices (affected by vulnerabilities disclosed at the NVD).\par

\begin{table*}[h!]
\caption{Description of dataset features}
\label{table:table1_dataset}
\centering
\begin{tabular}
{|p{2.5cm}            |p{2.1cm}               |p{1.5cm}           |p{5.6cm}|}
\hline
\textbf{Feature Name} &\textbf{Data Type} &\textbf{Unique Values} &\textbf{Details}\\
\hline 
Brand        &Categorical            &129     &Name of the device reported on the CVE.\\
\hline 
Product Type &Categorical            & 71     &Phrase describing the product.\\
\hline
Category     &Categorical            &  5     &SmartHome, Medical, Wearable, Telecomm, and Other.\\
\hline
Price        &Continuous             &Inf     &Reported in US Dollars.\\
\hline
Protocols    &Categorical            &8       &Protocols used in Communication Capability.\\
\hline
Data Storage &Binary                 &2       &Location of data Locally or Remotely.\\
\hline
Personal     &Binary                 &2       &Personal information data used: Yes\\
Information  &                       &        &or No.\\
\hline
Location Track &Binary               &2       &Tracks physical location: Yes or No.\\
\hline
Communication Capability&Categorical &31      &Communication technology.\\
\hline
Authorisation Encryption&Categorical &4       &Encryption used: Symmetric, Asymmetric, None, or Both.\\
\hline
Risk Score    &Categorical           &4       &From CVSS V3: Low, Medium, High, or  Critical.\\
\hline
\end{tabular}
\end{table*}

We carried a thorough analysis and justification of use to select a final set of features for the new dataset. For instance, the feature `Brand', can be used to identify brands of vulnerable products or rate brands by the number of vulnerable products. We also realised that the combination of this feature with others like `Type of Product' can help with the identification of the most vulnerable products amongst brands. We assessed all the twenty-seven initial features, agreeing on accepting only nine. Being these, Brand, Product Type, Category, Price, Protocol, Data Storage, Personal Information, Location Management, Communication Capability, Authorisation Encryption and Risk Score. Note Risk Score is used as the output feature. The full description of the dataset features is shown in Table \ref{table:table1_dataset}. The records added from the NVD range from the year 2013 to the year 2021. And, to keep the dataset updated with the latest records, carried a sweep to the NVD over the year 2021 up to June. Noteworthy to mention that as a part of our measurement mechanism on the behaviour of the models in the different classes, we added eight synthetic records, one record per classification belonging to two specific products (A smart speaker and a smart camera). We report 1,153 as the final number of records collected, in Table \ref{table:table_totDS} we disclose the distribution among classes.

\begin{table*}[h!]
\caption{Distribution of Dataset classes}
\label{table:table_totDS}
\centering
\begin{tabular}
{|p{3cm}            |p{2.1cm}               |p{2cm}|}
\hline
\textbf{Classification} &\textbf{Counts} &\textbf{Distribution \%}\\
\hline 
Low      &176        &15\%\\
\hline 
Medium   &138        &12\%\\
\hline 
High     &183        &16\%\\
\hline 
Critical &656        &57\%\\
\hline 
\end{tabular}
\end{table*}

\subsubsection{Data Imbalance}\label{subsub-section_DataImbalance}
The dataset is assessed to look for data imbalance. Note that whenever a dataset has a class imbalance, there exist strategies to reduce the imbalance and thus obtain better results. In the case of our dataset, we already know it contains imbalanced classes (Table \ref{table:table_totDS}), however, we still need to carry a cross-validation test to re-balance the parts of the dataset. The common approach to carry the cross-validation cannot be used in our dataset as it is known that this method is not recommended for a dataset with imbalance (It just splits the data into k folds). Rather we used the Repeated Stratified K-Fold Cross Validation (RSCV). The principal advantage of this approach is the method used to split the data in a distributed manner, where each fold contains the same percentage of samples per class, hence,  reducing the class imbalance.\par

\subsubsection{Data Transformation.}\label{subsub-section_Datatransformation}
We applied the Label Encoding technique that assigns a vector to each unique value, for each categorical feature. Later we applied the Normalised Text Frequency technique to convert the numbers into frequencies. To this point, the data on each feature has a representation in frequency. We found from this that the values are very sparse in some of the features.\par 

The sparseness of the data can be reduced through the application of a function. We tested four scaler functions and compared the data afterwards. We found the scaler that uses the Standard function as adequate to our data. This scaler reduced the gap of the peak values between classes. Note that leaving the data without being scaled harms the model (Comparative Results in section \ref{section4_Experiments}.\par

Although the data seems ready for the models, we applied another refinement function. This time, to organise the data into clusters (Utilises data similarity). For this, we used the K-means function in combination with another function that reduces the dimensionality (K-means does not perform well with high dimensional data). We compared two popular methods to carry this task are the PCA (Principal Component Analysis) and t-SNE (T-Distributed Stochastic Neighbour Embedding). The former generates clusters while also reduces dimensions by centring the data and measuring the distances from each point to the y-axis and x-axis, then the average distance for both axes is obtained, the data is centred through the obtained averages. The latter reduces the dimensionality of data with non-linear relationships, which is indeed the case with our data. T-SNE, unlike PCA, calculates a similarity measure, which is obtained by calculating the distance between points, finally, all the points with proportional similarities are clustered.  The effect of applying the t-SNE function into the data is that it clusters better the data, this due to its function that utilises similarity distance between points. However, this function demonstrated that rather than improving the model metrics, which affected contrarily, so we concluded to not using it. We show in table \ref{table:table3_othermetrics} the results of utilising the models with and without the t-SNE function. 
\subsubsection{Machine Learning Classification Models.}\label{subsub-section_Models}
We found it appropriate for our selected output label, based on the CVSS Severity, to use Classification Models. For this, we found that classification models work in individual format or an aggregated manner (ensemble mode) and that the latter has emerged as being better to tackle classification problems\cite{kotu}. Hence, we chose Ensemble models.\par

Ensemble models are designed to combine predictions from several built-in models, which in comparison to single models, can take advantage of the array capability of several sub-models to increase the prediction. Also, these models can improve the output. And lastly, their inner loss function helps to reduce the error and resolving bias problems from individual models.\par

From the various ensemble models for classification, we chose Gradient Boosting Decision Trees (GBDT) over AdaBoost Classification (ABC), Random Forest Classifier (RFC), Extra Tree Classifier (ETC), and Voting Classifier (VC). Note that in general, the listed models tend to manage differently their capability to reduce the prediction error. In particular, each model manages different approaches to improve the prediction. For instance, the ABC model manages the base models through weights and prediction repetition over the training dataset and returns the prediction with the least error rate, the RFC model builds a group of random decision trees containing random subsets of attributes. Here all the trees provide a predicted class value, and utilising the common voting approach they select the class with the majority of votes, the VC model is the simplest, which hosts multiple base models and the final output is the result of a majority voting process, and lastly, the GBC model combines several weak sum-models into a strong one, trains the base models in a sequence manner and assigns weights to every training record. The boosting process focuses on those records hard to classify by over-representing (assigning higher weights) them in the training set of the next iteration. Every new iteration on the model will focus on those records considered hard to classify. Furthermore, the results are combined from all the base learners through a majority voting approach.\par

We find the GBDT model with superior capabilities than the other models. One of the advantages of this model is the weight management strategy, which results in a better predictive approach (known as hard-to-classify). Furthermore, its sequential approach to order the internal models improves the final prediction. Thirdly, its built-in approach to managing class weights resolves imbalance problems. On the other hand, the most representative weaknesses we have with the voting approach (based on a simple majority of votes) is that it is not efficient in managing neutral cases\cite{cranor}. Added to this, the tuning process makes it difficult to find the best parameters and requires large searches. Finally, the process to reduce the loss may lead to overfitting outlier samples. Despite the intricacies of using the GBDT model we find it the best option amongst the selected models. Thus, we select GBDT as the primary model to use and for benchmark the RFC and VC. Note that we are discarding the ABC and ETC because in previous tests, these models returned the lowest performance metrics.\par

\subsubsection{Evaluation Metrics.}\label{subsub-section_Metrics}
The approaches we used to evaluate the selected classification models are through the application of the RSCV and the execution of the model after splitting the dataset into Training and Test. The results from the former approach will measure the accuracy from each model and based on the number of repetitions and k folds specified (2 repetitions and 5 folds in our case). For the latter approach, we will use the metrics Precision, Recall, F-1 and Accuracy. By itself, each metric provides part of the story, but together they help to confirm the correctness of the results from the selected model. Note that in multi-class prediction, Precision, Recall and F-1 have a variation in the calculation due to the classes \cite{grandini}, this is known at the Micro and Macro levels.
\begin{itemize}
\item \textbf{Micro Level}, computes the average of all samples without using class weights. This value is equal to the accuracy of the model. 
\item \textbf{Macro Level}, computes the overall mean of all classes, here the class weights are included.
\end{itemize}
Note that the Macro level approach for the metrics is recommended for datasets with imbalanced classes, which is the case of our dataset. Henceforth, the measurement of metrics will be focused on all the classes equally.\par

\section{Experiments and Results}\label{section4_Experiments}
\subsection{Experimental Setup}\label{sub-section4.2_ExperimentalSetup}
The validated dataset was uploaded to the Google Colab platform used for evaluations. Here, the resources are allocated as a default configuration with variable amounts such as 0.8-25 GB in RAM and 38-107 GB in Hard Drive space. The environment works under Python version 3.0 with the libraries Sklearn, Pandas, Numpy, Matplotlib, Seaborn, Pydrive, Phik, and Plotly.\par  

To tune the models' parameters, we used the tool `GridSearchCV'(from Scikit-learn\footnote{\url{https://scikit-learn.org/stable/}}) that carries a tuning process of the parameters with the most influence on the prediction, for each selected model. For instance, the tool provided that the adequate parameters for the GBC (Gradient Boosting Classifier) should be set with 10,000 as the number of estimators (number of boosting stages to perform), a learning rate of 0.01 (shrinks the contribution of each tree), as the maximum depth of 500 (of the individual regression estimators), and a minimum of impurity decrease value of 1e-2 (A node will be split based on this parameter). For the RFC, the tuning function recommended setting the parameter of class weight on `balanced' (To re-balance the classes by applying weights) and leaving the rest of the parameters with their default values. And, in the case of the VC (Voting Classifier) model, it receives the sub-classifier models AbC, GBC, ETC and RFC (each with the tuned parameters) and `type of voting' is set on `soft'. Note that all the parameters not mentioned, are left with their default values because of the models perform better with that configuration. In the next section, we will discuss the results from each model and carry a complete evaluation process.\par

\subsection{Evaluation Results}\label{sub-section4.3_EvaluationResults}

We show the results (Table \ref{table:table2_cv}) from running the selected models(GBC, RFC, and VC) through the RSCV and from not-applying/applying t-SNE clustering or PCA. The comparison is intended to show the negative effects in the accuracy metric. For better appreciation, the values equally or over 70\% are in bold.\par

\begin{table*}[h!]
\caption{Accuracy results from applying RSCV (5 folds, 2 repeats) over the selected models. Run-A refers to the results from the models without applying t-SNE clustering or PCA, Run-B refers to the results from the models after applying t-SNE clustering, and Run-C refers to the results from the models after applying PCA. The columns label represent the number of Run(1, 2) next to the number of fold(1-5), All the values are in \% format.}
\label{table:table2_cv}
\centering
\begin{tabular}{ 
|p{1.5cm}|p{.9cm}|p{.9cm}|p{1cm}|p{1cm}|p{1cm}|p{1cm}|p{1cm}|p{1cm}|p{1cm}|p{1cm}|p{1.2cm}|}
\hline
         &R1-F1 &R1-F2 &R1-F3 &R1-F4 &R1-F5 &R2-F1 &R2-F2 &R2-F3 &R2-F4 &R2-F5   \\
\hline 
Run-A& & & & & & & & & &  \\
\hline 
GBC      &67.1 &69.7 &69.3 &\textbf{70.9} &\textbf{72.2} &\textbf{73.6} &66.2 &66.7 &\textbf{70.9} &67.8\\
 \hline 
VC       &64.1 &65.8 &67.5 &67.8 &63.9 &\textbf{70.6} &67.1 &65.8 &67.0 &65.2 \\
\hline
RFC      &65.4 &68.0 &67.1 &\textbf{71.3} &65.7 &68.0 &66.7 &67.5 &67.8 &69.6  \\
\hline
Run-B& & & & & & & & & &  \\
\hline 
GBC      &58.4 &61.9 &66.2 &67.8 &64.3 &66.2 &67.5 &67.1 &64.8 &67.8 \\
\hline
VC       &60.2 &61.9 &66.7 &67.0 &63.9 &65.4 &68.8 &64.9 &64.3 &64.8  \\
\hline
RFC      &61.9 &67.5 &69.7 &69.6 &67.4 &67.5 &68.0 &\textbf{71.0} &67.0 &\textbf{72.6}  \\
\hline
Run-C& & & & & & & & & &  \\
\hline 
GBC      &62.0 &62.7 &66.4 &64.1 &63.3 &64.6 &65.3 &66.0 &61.9 &60.9 \\
\hline
VC       &62.3 &63.6 &63.2 &66.1 &59.6 &66.7 &67.1 &62.3 &61.3 &63.0  \\
\hline
RFC      &63.2 &66.7 &63.6 &64.8 &60.0 &65.8 &65.4 &64.5 &60.4 &67.0  \\
\hline
\end{tabular}
\end{table*}

In a second comparison we analysed the negative impact of using t-SNE clustering or PCA versus not using them. For this, we used the metrics Precision, Recall, and F-1, obtained from the testing results. In this analysis, we only show the results from the main model GBC. In Table  \ref{table:table3_othermetrics} we show the results of the comparison. Our quantitative benchmarks demonstrate that dimensionality reduction methods (PCS and t-SNE) reduce the performance of the selected models, with only three trials registering an average score above 70\% (shown in bold).\par

\begin{table*}[h!]
\centering
\caption{Precision, Recall and F-1, from applying GBC model. The section labelled WO/DR (without dimensionality reduction) shows the results from using the dataset without t-SNE clustering or PCA.  The section W/t-SNE (With T-SNE clustering applied) shows the results from using the dataset after applying t-SNE clustering. And the section W/PCA (With PCA applied) shows the results of using the dataset after applying PCA. The column labels represent the data classes Low, Medium, High, Critical, Macro(Macro Average Result) and Micro/ACC(Micro Average Result or Accuracy). All the values are in \% format.}

\begin{tabular}{||p{2cm}||p{1.42cm}|p{1.42cm}|p{1.42cm}|p{1.42cm}||p{1.42cm}||p{1.7cm}||}
\hline
\hline
\textbf{WO/DR}   &Low    &Medium    &High   &Critical   &Macro  &Micro/ACC\\
\hline
\hline
Precision       &\textbf{75.0} &50.0 &56.7 &\textbf{76.6} &64.6 &\textbf{71.0}\\
\hline
Recall          &63.2 &42.9 &53.1 &\textbf{83.5} &60.6 &\textbf{71.0}\\
\hline
F-1             &68.6 &46.2 &54.8 &\textbf{79.9} &62.4 &\textbf{71.0}\\
\hline
\hline
\textbf{W/t-SNE} &Low    &Medium   &High  &Critical   &Macro &Micro/ACC\\
\hline
\hline
Precision        &50.0 &59.1 &61.3 &\textbf{74.0} &61.1 &67.5\\
\hline 
Recall           &45.7 &46.4 &51.4 &\textbf{82.4} &56.5 &67.5\\
\hline
F-1              &47.8 &52.0 &55.9 &\textbf{78.0} &58.4 &67.5\\
\hline
\hline
\textbf{W/PCA} &Low    &Medium   &High  &Critical   &Macro &Micro/ACC\\
\hline
\hline
Precision        &69.2 &50.0 &59.4 &\textbf{73.8} &63.1 &68.8\\
 \hline 
Recall           &51.4 &42.9 &51.4 &\textbf{84.0} &57.4 &68.8\\
\hline
F-1              &59.0 &46.2 &55.1 &\textbf{78.6} &59.7 &68.8\\
\hline
\end{tabular}
\label{table:table3_othermetrics}
\end{table*}

\section{Discussion and Conclusion}\label{section5_Discussion}
\subsection{Discussion}\label{subsection_Discussion}

\begin{figure}
\includegraphics[width=0.8\textwidth]{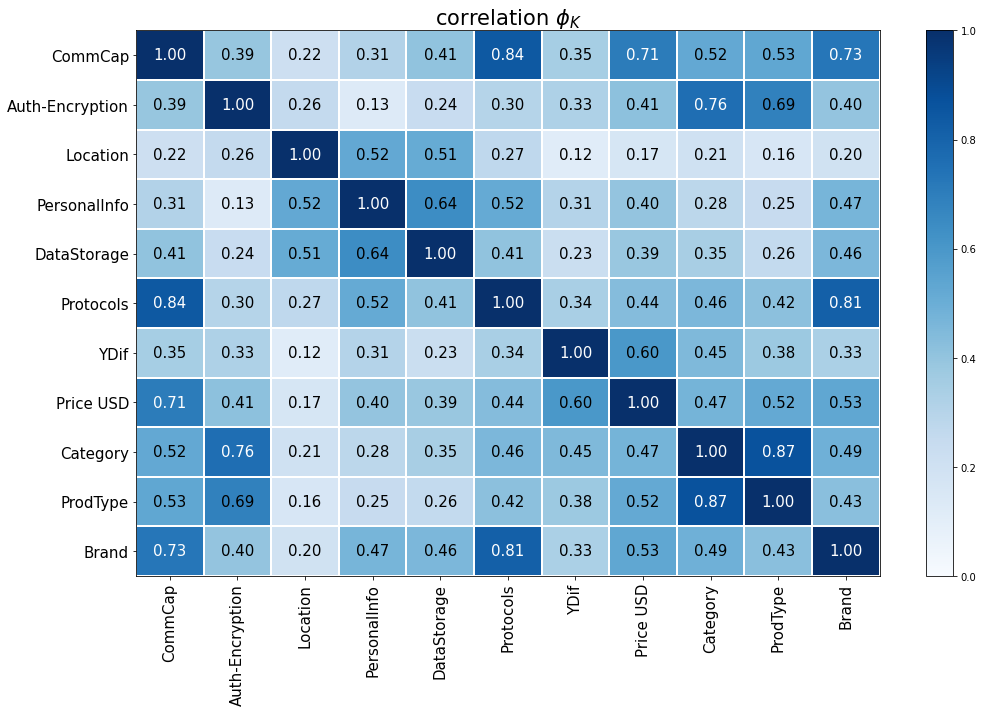}
\caption{Correlation plot for all the features in the dataset.}
\label{figure-correlation}
\end{figure}

From the dataset transformation process, we argue that our dataset features provided the models with the minimum features through which carried the prediction. To confirm this assumption, we used the correlation plot (see Figure \ref{figure-correlation}). Through the plot, we confirmed the correlation between pairs of features was not so high amongst most of the pair of classes, except for a couple of pairs that had values over 80\%. This means that one of the features may not be necessary to be included. For this, we applied a dimensionality reduction process where we drop those unnecessary features and check for variations in the metrics from the primary model. The results obtained allowed us to confirm that the dropped features are needed as every time one of those features were eliminated, the effect in the metrics was noticeable in a decrease of the values. We argue then that all the features from which we constructed our dataset are the minimum required for the model to provide decent prediction metrics. Another approaches to carry dimensionality reduction are t-SNE clustering and PCA. 
We applied both approaches with the purpose of improving the results for each model. For this, we tuned the relevant parameters and in some cases the data was clustered as a result. We observed in some other cases that these algorithms struggled to completely segregate points from the different classes. This could be responsible for lower model performance in comparison to the trials without the application of any dimensionality reduction. While this approach is effective with some problems, it was not effective with our risk prediction dataset. This was also the case when applied the PCA approach. 
\par

\subsection{Conclusion and Future Work}\label{subsection_Conclusion}
We have proposed a novel approach to predict the risk score in IoT devices. Our solution is complementary to IoT security threat detection and prevention mechanisms (e.g., \cite{lcncarlos}) and risk management frameworks. Our proposed risk score calculation solution enables enterprises to perform an easy and cost-efficient analysis of IoT devices to filter out the high-risk categories just by relying on their publicly accessible information. As part of this work, we created the first balanced vulnerability dataset for IoT devices with 1153 records accessible to future researchers in the field. The dataset was used to feed and train the GBDT machine learning model for the prediction of a risk score for IoT devices, which achieved 71\% prediction accuracy. In the future, we plan to extend this analysis and use the dataset to provide more specific predictions on the type of vulnerability of devices and suggest a set of recommended actions to the system administrators to protect their networks. \par

\bibliographystyle{splncs04}
\bibliography{bib}
\end{document}